# PCR-based isolation of multigene families: Lessons from the avian MHC class IIB

M. Promerová [1,2,3,*], R. Burri [1,4,5,*] & L. Fumagalli [1]

[1] Laboratory for Conservation Biology, Department of Ecology and Evolution, University of Lausanne, CH-1015, Switzerland

[2] Institute of Vertebrate Biology, Academy of Sciences of the Czech Republic, CZ-60365 Brno, Czech Republic

[3] Department of Medical Biochemistry and Microbiology, Biomedical Centre, Uppsala University, SE-75123, Sweden

[4] Department of Evolutionary Biology, Evolutionary Biology Centre, Uppsala University, SE-75236, Sweden

[5] Correspondence: Reto Burri, Tel. +46 (0)18 471 42 63; E-mail: reto.burri@ebc.uu.se

* These authors contributed equally to the work

## Abstract

The amount of sequence data available today highly facilitates the access to genes from many gene families. Universal primers amplifying the desired genes over a range of species are readily obtained by aligning conserved gene regions, and laborious gene isolation procedures can often be replaced by quicker PCR-based approaches. However, in case of multigene families, PCR-based approaches bear the risk of incomplete isolation of family members. This problem is most prominent in gene families with highly variable and thus unpredictable number of gene copies among species, such as in the major histocompatibility complex (MHC). In the present study we (i) report new primers for the isolation of the MHC class IIB (MHCIIB) gene family in birds, and (ii) share our experience with isolating MHCIIB genes from an unprecedented number of avian species from all over the avian phylogeny. We report important and usually underappreciated problems encountered during PCR-based multigene family isolation, and provide a collection of measures that may help to significantly improve the chance of successfully isolating complete multigene families using PCR-based approaches.

# Introduction

Genes of the major histocompatibility complex (MHC) family have become one of the most important model systems for the study of adaptive genetic diversity and host-parasite coevolution (Piertney& Oliver 2006; Sommer 2005), and have been instrumental for the understanding of multigene family evolution (e.g. Nei& Rooney 2005). While for many genes the development of primers and protocols to amplify desired gene regions in a broad range of species is a standard and feasible task, the isolation and characterization of MHC genes in multiple species has proven notoriously challenging due to their complex evolution. The pronounced molecular and evolutionary dynamics of MHC genes are governed by high rates of gene duplication, frequent inter- and intragenic recombination and gene conversion, and strong balancing selection is usually acting on these genes (Klein *et al.* 1993). The simultaneous action of these evolutionary forces often causes rapid divergence among MHC sequences of even closely related species. Consequently, conserved regions which could serve as PCR priming sites over a broad range of species are accordingly scarce and difficult to identify.

In birds, these problems appear to be even more pronounced than in other taxa. Elevated rates of concerted evolution homogenize duplicates within species and enhance the divergent evolution of genes among species (Hess& Edwards 2002). Primers amplifying avian MHC so far were highly degenerate and amplified only short fragments (Edwards *et al.* 1995), or were usually developed specifically for every species or phylogenetic group of interest. This task often includes laborious procedures such as genome walking, RACE, or genomic library screening, implying a lengthy isolation procedure. Although in the past years several studies reported successful cross-species amplification of rather long sequences of avian MHC class IIB genes in related species (e.g. Alcaide *et al.* 2007; Burri *et al.* 2008a), there is no consensus on universal primers and protocols that can be used for the isolation of MHCIIB genes in a broad range of bird taxa. The lack of generally applicable molecular markers for the study of MHC genes in birds has not only lead to an underrepresentation of studies on avian MHC evolutionary ecology compared to other taxonomic groups, but also to a so far only rudimentary understanding of the long-term evolutionary history of the avian MHC. While for mammals studies documenting the relationships of MHC genes among species have been published already in the early nineties (Nei & Hughes 1992), first such studies from birds were reported only recently (Burri *et al.* 2008a; Burri *et al.* 2010). Thus, the development of universal molecular tools facilitating the isolation of avian MHC genes is an important contribution to advance both avian MHC evolutionary ecology and the understanding of the long-term evolutionary history of the avian MHC.

For the reasons pointed out above, primers and protocols that are applicable to each and every species will presumably remain a mere vision. However, the amount of MHC sequence data in genetic databases has greatly increased over the past decade, and promises to significantly enhance the design of fairly universally applicable PCR primers. While PCR-based isolation approaches are straightforward compared to other laborious procedures, especially in multigene families like MHC they bear the risk of missing some genes or entire gene lineages (Kanagawa 2003; Wagner *et al.* 1994). This is because (i) primers may be specific to only a subset of genes within a species, (ii) molecular properties may differ between genes, or (iii) random processes during PCR can result in biased amplification (Wagner *et al.* 1994). Avoidance of such biases during isolation of MHC genes is extremely important in order to capture the complete MHC diversity in the context of evolutionary ecological studies, and it is also essential for studies of evolutionary history of these genes. Incomplete isolation of part of the gene family, or in the worst case whole gene lineages, can result in false interpretation as gene loss, or in the overestimation of rates of concerted evolution (see discussion in Burri *et al.* 2010).

In the present study we report a new set of PCR primers and advice for the isolation of the MHCIIB gene family in birds based on our extensive experience gained during the isolation of MHCIIB genes in an unprecedented number of species belonging to a number of avian orders. Our aim was twofold, notably (i) to target the widest possible phylogenetic range of species, and (ii) to minimize the risk of missing genes and thus to improve the chance of successfully isolating the complete MHCIIB family in birds. Moreover, we report the most common and often underappreciated or unexpected problems with PCR-based multigene family isolation that should assist researchers who are planning to isolate the MHCIIB gene family in birds in particular and multigene families in general.

## Material & Methods

*DNA extraction*

DNA from 37 bird species was extracted from fresh muscle tissue (*S. camelus*, *A. atthis*), muscle tissue stored in 96% ethanol (*C. casuarius*, *D. noveahollandiae*, *R. aquaticus*), blood stored in EDTA buffer or 96% ethanol (*C. livia*, *A. melba*, all Procellariformes, Charadriiformes and Strigiformes, *J. torquilla*, *P. socius*), or feathers (*D. major*, *T. hartlaubi*, all Ardeidae and Accipitridae) using the DNeasy blood and tissue kit (Qiagen, Hombrechtikon, Switzerland) following the manufacturer's protocol.

*Primer design for MHCIIB amplification*

In order to identify conserved regions which can be used for priming of the longest possible MHCIIB sequences, we aligned MHCIIB exon 1 and exon 4 sequences, respectively, of all bird species for which they were available on GenBank by July 2010. Based on these alignments we developed two slightly degenerated forward primers lying in the same conserved exon 1 region as Ekblom *et al.*'s (2003) primer 34F (F1, F2; **Table 1**) and two reverse primers (R1, R2; **Table 1**) situated in exon 4. These were used for PCR amplification in different combinations and with different PCR conditions according to the criteria outlined below. The expected length of the avian MHCIIB sequence spanning from exon 1 to exon 4 can be highly variable due to strongly varying intron lengths among species. The minimal expected length of fragments amplified with our primers is ~1.3 kb (e.g. ostrich, *S. camelus*), but the fragment length can reach up to >2.5 kb in species like passerines (Passeriformes) (e.g. Edwards *et al.* 2000), parrots (Psittaciformes) (Hughes *et al.* 2008), and falcons (Falconidae) (Alcaide *et al.* 2007). Different primer combinations were tested where necessary. **Table 2** reports for each species the primer combination(s) which provided the whole set of amplified alleles, and in parentheses the ones which worked but amplified only a subset of alleles. In species where our primers amplified a lower than subjectively expected number of alleles, we tested Alcaide's AlEx3R reverse primer in exon 3 (Alcaide *et al.* 2007) situated in the presumably most conserved region of avian MHCIIB in combination with our forward primers. For palaeognath species, we developed primers based on the alignment of our first obtained sequences (isolated with F1/R1 and F2/R1, **Table 2**) with little spotted kiwi (*Apteryx owenii*) sequences (Miller *et al.* 2011).

*PCR, cloning, and sequencing*

PCR amplification of MHCIIB genes poses a number of challenges, and several properties specific to these genes have to be taken into account. First, avian MHCIIB genomic sequences are extraordinarily GC-rich, with a GC-content of usually >70% (in some introns even 80%).

Consequently, nucleotide bonds between strands are very strong and denaturation of the MHC region can be difficult. Moreover, in some species MHCIIB genes contain long homopolymers and repetitive sequences in the introns that further complicate amplification. E.g. grey partridge (*Perdix perdix*) has a long GGCCCC hexanucleotide microsatellite in intron 1 and its reverse complement CCGGGG in intron 3. These appear to form an extremely strongly zipped hairpin, which makes amplification impossible if the DNA is not adequately denatured (Promerová & Bryjová, unpublished). Therefore, an initial PCR step of thorough denaturation and PCR additives modifying the DNA's melting behaviour and facilitating amplification are important (e.g. Qiagen's Q Solution or DMSO). A hot start DNA polymerase that supports long initial denaturation without losing amplification efficiency may be required. Finally, amplification should preferably be specific to MHCIIB genes, but without being specific to only a subset of MHCIIB genes and thus missing others.

The standard PCR conditions taking into account these requirements and considering the expected length of the amplicon for all species were as follows. PCR was carried out on a Biometra T1 gradient thermocycler in a final volume of 25 µl containing 1x buffer Gold, 2 mM $MgCl_2$, 0.2 mM dNTP, 0.5 µM of each forward and reverse primer, 1x Q Solution (Qiagen, Hombrechtikon, Switzerland), 1 U of AmpliTaq Gold (Applied Biosystems, Rotkreuz, Switzerland) and 2 µl of extracted DNA (DNA quality was assessed by gel electrophoresis). PCR cycling included initial denaturation for 10 minutes at 95°C, 35 cycles consisting of denaturation at 95°C (40 sec), annealing at 63°C, 65°C and 68°C (40 sec), and elongation at 72°C (2 min). A final step at 72°C for 7 min was introduced to complete extension. PCR products were stained with ethidium bromide and visualized by agarose (1.5%) gel electrophoresis. Amplicons were purified using the Wizard SV Gel and PCR Clean-up System (Promega, Dübendorf, Switzerland). If necessary, candidate bands were excised from gel and purified using the MinElute Gel Extraction Kit (Qiagen). PCR products were then cloned using the pGEM-T Easy Vector System (Promega). For each cloned PCR product a minimum of 6-12 clones were sequenced usually from extracted plasmids using the T7 and SP6 vector primers and in a few cases directly from colony PCR. Products which could not be completely sequenced (usually due to homopolymers and repeats, or because the amplified fragments were too long), were subjected to difficult-template sequencing with Macrogen (Seoul, South Korea). Standard Sanger sequencing was conducted at Microsynth's (Balgach, Switzerland) or Macrogen's (Seoul, South Korea) facilities. Except for a few species for which we did not pursue the isolation (**Table 2**), we conducted a minimum of two (usually 3) independent PCRs and performed the above described procedure for each PCR product separately. Sequences were considered confirmed if they were obtained in at least two independent manipulations. Highly divergent sequences which were not confirmed but excluded to be contaminants or recombinants, and highly divergent from other sequences are included in the 'divergent sequences' count (**Table 2**). Sequences will be published elsewhere.

## Results & Discussion

We successfully isolated MHCIIB sequences in 37 species from 13 avian orders/families from all over the avian phylogeny (**Figure 1**, **Table 2**), demonstrating that the reported primers may be expected to work for MHCIIB isolation in a broad range of bird taxa. For 30 of these species we confirmed the sequenced alleles by at least two to three independent PCRs, and thus obtained estimates of the minimal number of alleles and loci. The number of confirmed alleles per species ranged from one to five, implying that the species possess at least one to three MHCIIB loci. If divergent but not confirmed sequences were included we found up to seven alleles per species, suggesting that these species possess at least four loci. While we think that these estimates are realistic for most species, in passerines we expect more loci than the two to three identified, because MHCIIB is known to be highly duplicated in this order. The underappreciated problem here is that confirmation of alleles is a significantly more difficult task in species with a highly duplicated MHC: First, the probability of observing a given allele repeatedly is significantly lower. Second, with increasing number of loci the number of possible artifactual recombinants arising during PCR increases drastically. Patterns consistent with the latter were observed in azure-winged magpie (*Cyanopica cyana*) and in both species of storm petrels (*Oceanodroma*). We indeed expect the actual number of loci in these species to be above the estimates provided in **Table 2**.

*General amplification success and amplification patterns*

Our primers yielded MHCIIB-positive bands at the expected lengths in 37 species (**Table 2**). Primer pair F2/R1 was usually tested first due to the primers' ideal thermodynamic properties, and successfully amplified MHCIIB in most cases. Primer combinations F1/R1 and/or F2/R2 were tested successfully in most species where the previous primer pair was not successful. F1/R2 was tested only if all previous primer combinations failed. In general, for many if not most species, we were able to specifically amplify first MHCIIB bands in less than two days of work. Compared to several months of establishment needed for instance for the barn owl (*Tyto alba*) MHCIIB a few years ago (Burri *et al.* 2008b) this appears to be a significant improvement.

For a considerable number of species, the amplification resulted in multiple (usually two to three) fragments within the expected length range. In most of these species, indeed all of the different fragments proved to be MHCIIB sequences of diverse length (**Figure 2**,

*Pachyptila desolata, Ciconia ciconia, Limosa limosa*). Weak, unspecific bands outside the expected length range usually disappeared when the number of PCR cycles was reduced to 32 and/or when using the more specific Biometra T Professional thermocyclers (a machine with a high ramp speed). However, in a few species additional non-MHCIIB bands could not be eliminated; especially around 0.9 kb and >3 kb (see **Figure 2**, *Ciconia ciconia, Cyanopica cyana*). Where necessary, we excised the fragments of expected length from gel (note that for the species shown in **Figure 2** this was not necessary, because the unspecific bands are much weaker and are not ligated during cloning).

*Cases of failed MHCIIB amplification*

In species belonging to three avian orders/families amplification of MHCIIB failed completely, namely in falcons (*Falco tinnunculus*; Falconidae), parrots (*Agapornis lilianae*, *Cacatua galenta*, *C. leadbeateri*, *C. triton*, *Psittacus erythacus*; Psittaciformes), and penguins (*Aptenodytes patagonicus*, *Eudyptes chrysocome*, *E. chrysolophus*, *Pygoscelis papua*; Sphenisciformes). Amplification also failed for museum samples (muscle in 96% ethanol) of three tested suboscine passerines (*Muscivora tyrannus*, *Pipra fasciicauda*, *Pitta elliotii*) despite considerable effort in the latter two, although we successfully amplified MHCIIB in two oscine passerine species, the azure-winged magpie (*Cyanopica cyana*), and the sociable weaver (*Philetairus socius*). Published sequences of falcons, parrots and passerines all display extraordinarily long introns (Alcaide *et al.* 2007; Edwards *et al.* 2000; Hughes *et al.* 2008), which often contain highly repetitive sequences. These may indeed inhibit amplification of long sequences. Moreover, the qualities and/or quantities of DNA in the samples from all these species but the falcon were rather poor. Thus, most probably sequence length and DNA quality sign responsible for the amplification failure in these species. Long sequences presumably need other protocols not tested in the framework of the present study (e.g. amplification in overlapping shorter fragments, these ideally including the highly variable exon 2 in order to enable phasing of the separate fragments).

Still we cannot exclude that our primers are not applicable in a number of species. Indeed, any primer may mismatch any species in a random manner, such as exemplified by Bulwer's petrel (*Bulweria bulwerii*). In this species both forward primers work with either R2 or AlEx3R, but not with R1, although the latter is the reverse primer of choice in most other Procellariformes (**Table 2**). Similarly, both reverse primers failed to amplify Razorbill (*Alca torda*), while AlEx3R was successful with either of the forward primers, and R1 and/or R2 working perfectly in the other Charadriiformes (**Table 2**).

*Primer-dependant PCR-biases*

The choice of primers can have a crucial effect on how many and which genes are amplified in a given species. Although primer combinations that worked well in one species often did so also in related species, this prediction does not always hold (see examples from Procellariformes and Charadriiformes above). In the less problematic of cases amplification simply fails. A much larger problem arises if PCR is biased in favour of particular genes or alleles, because such a bias easily remains undetected when not considered. Therefore, and as illustrated by the following examples, we strongly encourage researchers to test a number of primer combinations, in order to reduce the risk of primer-dependant PCR-biases. We observed such cases in two of the three palaeognath species (*Casuarius casuarius*, *Dromaius novaehollandiae*), in white stork (*Ciconia ciconia*), and in thin-billed prion (*Pachyptila belcheri*). In these species additional bands appeared only in later PCR tests, when using a number of primer combinations. Similarly, new sequences of equal length to previously isolated ones appeared in Cory's shearwater (*Calonectris diomedea*) and Monteiro's storm petrel (*Oceanodroma monteiroi*) when different reverse and forward primers, respectively, were applied. As exemplified by the latter cases, it appears crucial to not only judge amplification success between primer combinations according to gel electrophoresis. In order to maximise the chance to isolate all target genes in a given species, without exception amplicons from several (ideally all) successful primer combinations should be cloned and sequenced.

Finally, we note that in some species all tested primer pairs yielded highly biased PCR products. We observed such a case in the pigeon (*Columba livia*), where out of 70 sequenced clones 67 were from a first and only three from a second confirmed sequence. In white stork (*Ciconia ciconia*) three of the five confirmed sequences together made up only seven clones, while the other two were sequenced in 12 and 20 clones, respectively. Thus sequencing of a moderate amount of clones from several primer combinations might in some cases not be sufficient to capture all alleles, but a high cloning and sequencing effort is necessary during the initial isolation procedure.

*Annealing temperature-induced PCR-biases*

Due to the high GC-content of avian MHCIIB genes, annealing temperatures ($T_a$) as high as 68°C usually worked best for all reported primers. We therefore usually tested only $T_a$ from 63°C upwards. In many species with MHCIIB bands at different lengths, all temperatures worked similarly well as judged from gel. However, for some of them the amplification patterns strongly varied depending on temperature. This is exemplified by the patterns from common guillemot (*Uria aalge*). In this species even a $T_a$ difference of 2-3°C caused preferential amplification of the longer products, and excision from gel and sequencing of the bands showed that both were MHCIIB (not shown).

In a few other species, we observed temperature-dependant variation in amplification patterns which are not intuitive. In general, PCR amplification is expected to be more specific at higher $T_a$. Contrary to this, in several species certain MHCIIB bands were exclusively or better amplified at high $T_a$. We suppose that this is yet another effect of high GC-content, causing the double-strand to remain denatured during annealing and/or elongation only at high $T_a$. For obvious reasons, this pattern was detected on gel only in species possessing MHCIIB bands of different lengths. Researchers should be aware that such temperature-dependant biases may equally occur in other species, though might be more difficult if not impossible to observe due to the MHCIIB bands' equal length. In such cases this temperature-dependant PCR-bias can again only be detected by sequencing a sufficient number of clones from PCRs performed at different $T_a$.

Though we argue why high $T_a$ may be needed to isolate avian MHCIIB genes, it is again important to note that PCR conditions should be kept only as specific as necessary in order to amplify all the genes present. As outlined above, high $T_a$ is not necessarily more specific when amplifying avian MHCIIB, but different $T_a$ may cause different sequences to be amplified. We therefore recommend sequencing products from several $T_a$. In many cases, this can be achieved by direct sequencing of the products and a careful check of double peaks. Since insertions and deletions are usually situated in the 5'-region of the genes, this approach also works in species with bands at different sizes (as long as all of them are MHCIIB) if the reverse primer is used for sequencing (we recommend AlEx3R).

*Between-individual variation in amplification patterns*

The genes of the MHCIIB are amongst the most polymorphic in vertebrates, and in many species most individuals exhibit a different MHC genotype. If variation also segregates at the priming sites that we use here to isolate MHCIIB, amplification success and patterns can be expected to vary between individuals. Although our priming sites are highly conserved between species, we tested for within-species/between-individual variation in amplification patterns by amplifying a number of individuals in two species, namely blue petrel (*Halobaena caerulea*) and bar-tailed godwit (*Limosa limosa*). Both species exhibit two MHCIIB bands of different length, which usually amplify at similar strength as judged from agarose gels (see **Figure 2** for bar-tailed godwit; blue petrel shows a very similar pattern). In blue petrel, in one out of ten individuals the short band amplified considerably weaker than the long one, and in two petrels the short band was almost completely absent (the fact that the short band is affected speaking against problems with DNA quality here). In bar-tailed godwit, out of seven individuals two displayed a much stronger short band, and in one individual only the short band was amplified.

These results indicate that within species amplification patterns can differ significantly among individuals, with some individuals showing much stronger PCR-bias than others. Especially researchers planning to genotype MHCIIB in different populations should pay significant attention to between-individual variation already during the isolation procedure. Although for a number of reasons it is generally advisable to base the isolation of genes on a single reference individual, we recommend including a handful of individuals from as broad a geographic range as possible in PCR tests preceding extensive cloning. In species exhibiting a single band, several individuals should be cloned and sequenced. In case of both variably and constantly strong PCR-bias between genes, we advise to design new species-specific primers based on the first obtained sequences and alignments with related taxa, such as performed for palaeognath species in the present study.

## Conclusions

The reported primers and protocols provide easy and quick access to MHCIIB genes in a wide range of bird species. **Figure 3** summarizes some critical advice for the PCR-based isolation of multigene families based on our experience with the isolation of MHCIIB genes in birds. As a general rule it is advisable to clone and sequence products from several primer combinations, from several annealing temperatures and from several individuals in order to capture the whole range of MHCIIB genes in a target species. With respect to the sometimes considerable cloning effort, clonal high-throughput sequencers may open promising perspectives.

The sequences which can be obtained using our primers and conditions include the whole peptide-binding region, which is usually targeted by MHC evolutionary ecology studies. With the recommended protocols also sequences of the flanking introns and exon 3 are usually easily obtained, which in many species hold gene- or gene lineage-specific sites that can be further used to specifically amplify single genes or gene lineages. By assisting the development of specific amplification of single genes, or at least reducing the complexity of multigene amplicons in a number of avian species, the reported primers and protocols will significantly increase the power of future studies in MHC evolutionary ecology. Finally, the possibility to isolate MHCIIB genes all over the avian phylogeny opens extraordinary perspectives for comparative studies to further the understanding of the evolutionary history of avian MHCIIB and avian pathogen resistance.

## Acknowledgements

We thank Francesco Bonadonna and Maria Strandh, Céline Serbielle and Karen McKoy, the Natural History Museum Stockholm, Michel Gauthier-Clerc, Joël Bried, Raphael Arlettaz, Theunis Piersma and Yvonne Verkuil, the Zoo La Garenne, and Glenn Yannic, who kindly provided tissue and DNA samples. Robert Ekblom and Andrea Šimková provided precious comments on an earlier version of the manuscript. Thanks to Hilary Miller for providing sequence information from little spotted kiwi prior to online publication. This study was supported by the Swiss National Science Foundation grant 3100A0-109852/1 to LF. MP was supported by the Czech Science Foundation (project no. P505/10/1871).

## Figure Legends

**Figure 1** – Phylogeny of avian orders/families indicating the phylogenetic positions of successful MHCIIB isolation. The topology follows Hackett *et al*. (2008).

**Figure 2** – Range of banding patterns of successful MHCIIB amplification in species with specific amplification of MHCIIB, which were directly cloned.

**Figure 3** – Important points to consider during the isolation of the avian MHCIIB multigene family.

## Tables

**Table 1 –** PCR primers and nested sequencing primers.

| Primer | Abbreviation | Sequence | Position | Purpose |
|---|---|---|---|---|
| AvesEx1-F1 | F1 | 5' - ACTGGTGGCACTGGTGG**Y**GC - 3' | exon 1 | isolation |
| AvesEx1-F2 | F2 | 5' - GCACTGGTGG**Y**GCTGGGAGC - 3' | exon 1 | isolation |
| AvesEx4-R1 | R1 | 5' - GAGCCCCAGCGCCAGGAAG - 3' | exon 4 | isolation |
| AvesEx4-R2 | R2 | 5' - GAAGA**Y**GAG**B**CCCAGCACGAAGC - 3' | exon 4 | isolation |
| AlEx3R [1] | - | 5' - CACCAGCA**S**CTGGTA**S**GTCCAGTC - 3' | exon 3 | isolation |
| RatiteEx1-F3 | F3 | 5' - GGCTCAGCTCACCTGT**S**GTCTCC - 3' | exon 1 | isolation Palaeognathae |
| RatiteEx1-F4 | F4 | 5' - GCTCACCTG**WS**GTCTCCT**Y**GCC - 3' | exon 1 | isolation Palaeognathae |

[1] Alcaide *et al.* (2007)

440 **Table 2** – PCR conditions and results for successfully amplified species.

| Order / Family | | Species | | Primers | $T_a$ (°C) | Excised | No. Confirmed Alleles | No. Divergent Alleles [1] | No. loci |
|---|---|---|---|---|---|---|---|---|---|
| Struthioniformes | | *Casuarius casuarius* | Southern Cassowary | F3/AlEx3R (F1/R1) | 60 (65) | No | 3 | 3 | 2 |
| | | *Dromaius novaehollandiae* | Emu | F3/AlEx3R (F1/R1) | 60 (62-63, 60) | No | 2 | 3 | >1 |
| | | *Struthio camelus* | Ostrich | F2/R1, F1/R1, F1 to F4/AlEx3R | 63-65, 65, 60-63 | No | 2 | 2 | 1 |
| Columbiformes | | *Columba livia* | Feral Pigeon | F2/R1 | 68 | No | 2 | 3 | 2 |
| Apodiformes | | *Apus melba* | Alpine Swift | F2/R1, F1/R1 | 65 [4], 63 | Yes | 3 | 3 | 2 |
| Gruiformes | | *Rallus aquaticus* | Water Rail | F2/R1 | 61.2 [5] | Yes | 4 | 5 | >2 |
| Musophagiformes | | *Tauraco hartlaubi* | Hartlaub's Turaco | F2/R1 | 68 | No | 4 | 4 | 2 |
| Procellariiformes | | *Bulweria bulwerii* | Bulwer's Petrel | F2/R1 | 68 | No | - | 4 | - |
| | | *Calonectris diomedea* | Cory's Shearwater | F1/AlEx3R (F2/R2, F1/R1) | 68; 63 | No | 2 | 5 [9] | >2 |
| | | *Diomedea exulans* | Wandering Albatross | F2/R1, F1/R1 | 68 | No | 3 | 4 | 2 |
| | | *Halobaena caerulea* | Blue Petrel | F2/R1 | 68 | No | 3 [8] | 4 | 2 |
| | | *Oceanodroma castro* | Madeiran Strom Petrel | F2/R1, F1/R1 | 68 | No | 5 | 7 | >3 |
| | | *Oceanodroma monteiro* | Monteiro's Storm Petrel | F1/R1 (F2/R1) | 68 | No | 5 [8] | 7 | >3 |
| | | *Pachyptila belcheri* | Thin-billed Prion | F1/AlEx3R (F2/R1) | 65-68 (68) | No | 5 | 6 | 3 |
| | | *Pachyptila desolata* | Antarctic Prion | F2/R1 | 68 | No | 4 | 5 [10] | 3 [10] |
| | | *Pagodroma nivea* | Snow Petrel | F2/R1 | 68 | No | 3 [8] | 4 | 2 |
| | | *Procellaria aequinoctialis* | White-chinned Petrel | F2/R1 | 68 | No | 4 [8] | 5 | >2 |
| | | *Puffinus baroli* [2] | Little Shearwater | F2/R1 | 68 | No | - | 2 | - |

441 **Table 2** – continued.

| Order / Family | | Species | Primers | $T_a$ (°C) | Excised? | No. Confirmed Alleles | No. Divergent Alleles [1] | No. loci |
|---|---|---|---|---|---|---|---|---|
| Ciconiiformes | *Ardea cinerea* | Grey Heron | F2 / R1, F1 / R1 | 68, 65 | No | 2 | 2 | 1 |
| | *Bubulcus ibis* | Cattle Egret | F2 / R1, F1 / R1 | 68, 65 | No | 2 | 3 | >1 |
| | *Ciconia ciconia* | White Stork | F1 / R1 (F2 / R1) | 68 | No | 5 | 6 | 3 |
| Charadriiformes | *Limosa limosa* | Black-tailed Godwit | F2 / R2 | 68 | No | 4 | 4 | 2 |
| | *Philomachus pugnax* | Ruff | F2 / R1 | 68 | No | 5 | 6 | 3 |
| | *Pagophila eburnea* | Ivory Gull | F2 / R1 | 68 | No | 5 | 5 | 3 |
| | *Rissa tridactyla* [3] | Black-legged Kittiwake | F2 / R1 | 65-68 | No | - | - | - |
| | *Alca torda* [3] | Razorbill | F2 / R1 | 63 | No | - | - | - |
| | *Fratercula arctica* [3] | Atlantic puffin | F2 / R1 | 63-68 | No | - | - | - |
| | *Uria aalge* [3] | Common Guillemot | F1 / AlEx3 | 63-65 | No | - | - | - |
| Accipitridae | *Buteo buteo* | Common Buzzard | F2 / R1, F1 / R1, F1 / AlEx3R | 65 [6], 63 [6], 58-68 | Yes | 1 | 3 | >1 |
| | *Milvus milvus* | Red Kite | F2 / R1, F1 / R1, F1 / AlEx3R, F2 / AlEx3R | 63 [6], 68 [6], 58-68 | Yes | 5 | 5 | 3 |
| Strigiformes | *Strix aluco* | Tawny Owl | F2 / R2 | 65 [7] | No | 5 | >5 | >3 |
| | *Tyto alba* | Barn Owl | F2 / R2 | 68 [6] | No | 3 | 4 | 2 |
| Piciformes | *Dendrocopus major* | Greater Spotted Woodpecker | F2 / R1 | 68 | No | 2 | 2 | 1 |
| | *Jynx torquilla* [2] | Wryneck | F2 / R1 | 68 | Yes | - | 1 | - |
| Coraciiformes | *Alcedo atthis* | Common Kingfisher | F1 / R1, F2 / R1 | 68 | No | 3 | 3 | 2 |
| Passeriformes | *Cyanopica cyana* | Azure-winged Magpie | F1 / R1 | 68 | No | 3 | 6 | >2 |
| | *Philetairus socius* [2] | Sociable Weaver | F2 / R1 | 65 | No | - | 1 | - |

442

443  Primer names correspond to abbreviations indicated in Table 1. Primer combinations in brackets work but amplify only a subset of sequences. Subsequent conditions if not
444  indicated differently are valid for all primer combinations. Else the order of conditions corresponds to the order of primer combinations.
445  [1] Divergent alleles include alleles which could not be confirmed but are as divergent as to be considered real. They do not include recombinant sequences.
446  [2] Sequences were isolated from these species, but isolation was not pursued.
447  [3] These species have not been cloned. MHCIIB-identity was confirmed by direct sequencing of the products using the reverse PCR-primer.
448  [4] 6 min denaturation; 35 cycles with 30 sec denaturation, 30 sec annealing and 1 min elongation. 0.6 U Qiagen Taq polymerase.
449  [5] 6 min denaturation; 35 cycles with 35 sec denaturation, 30 sec annealing and 1 min elongation. Qiagen Taq polymerase.
450  [6] Standard protocol with 32 cycles.
451  [7] Touch-down PCR: 6 min denaturation, 10 cycles à 30 sec denaturation, 30 sec annealing (68-65), 1 min elongation; 20 cycles with annealing at 65°C. Primers 0.4 uM. 0.6
452  U Taq polymerase.
453  [8] Banding pattern confirmed with primer AlEx3R.
454  [9] Number takes into account that the directly sequenced PCR product (not cloned) that unambiguously indicates a further highly divergent sequence.
455  [10] Number takes into account that the longest band like in *P. belcheri* is also present in this species in PCRs using the AlEx3 primer, but cloning failed.

456 # Figures

457 **Figure 1**

458

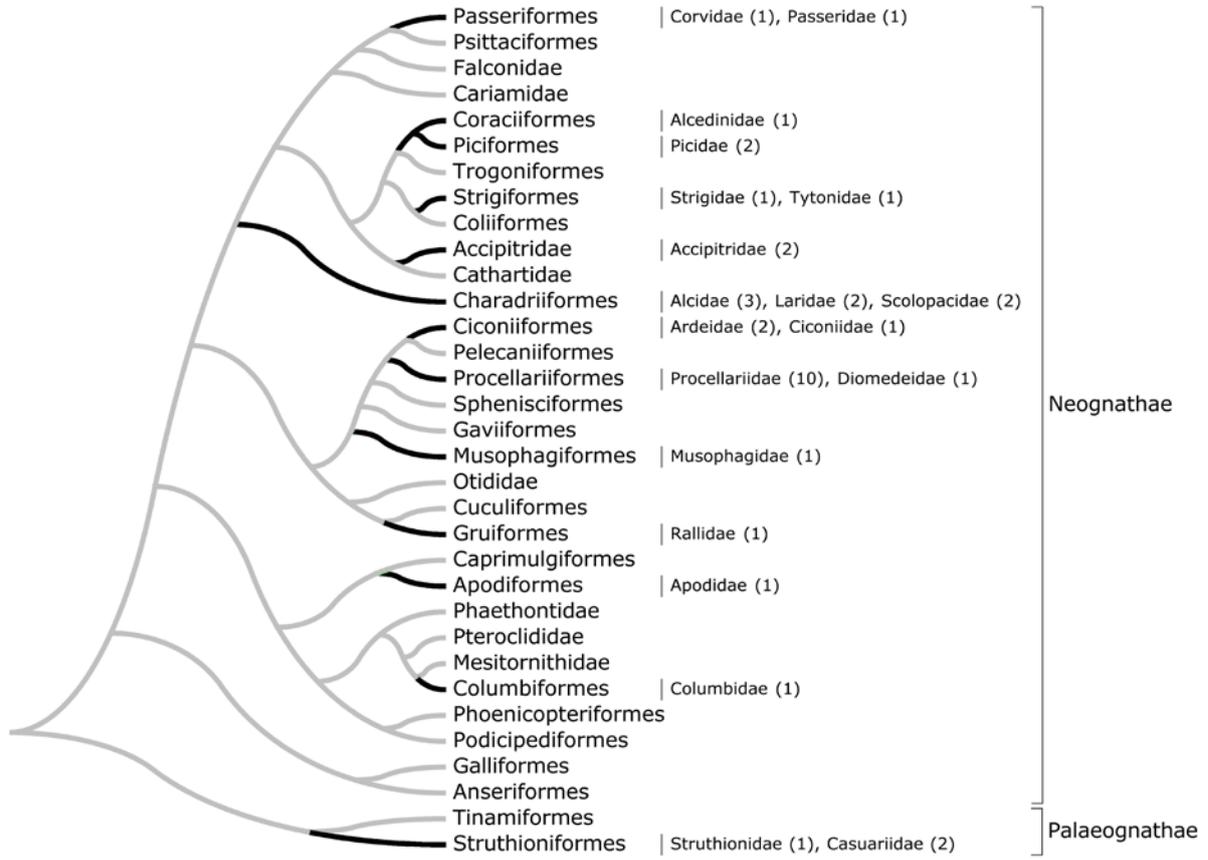

459
460

**Figure 2**

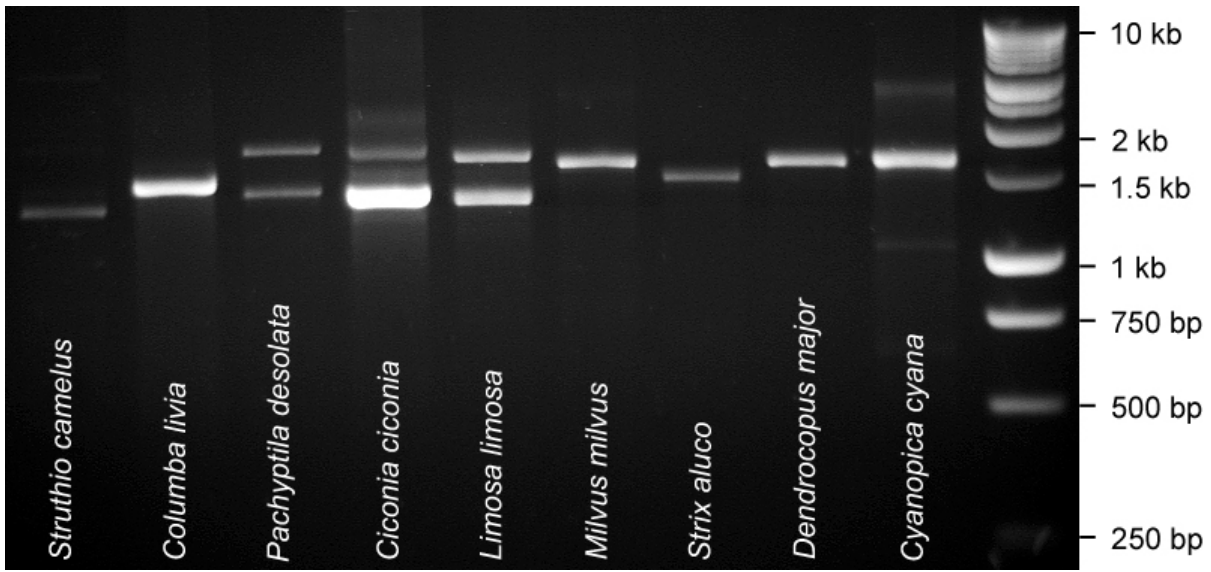

**Figure 3**

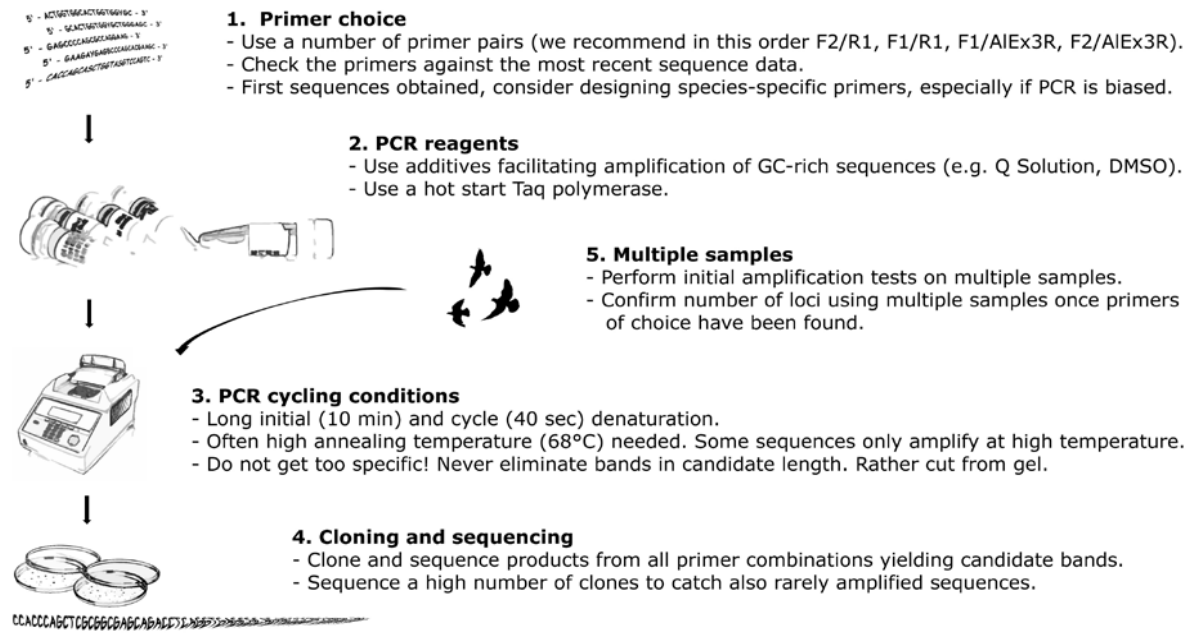